\documentclass[aps,prl,twocolumn,groupedaddress,showpacs]{revtex4}
\usepackage{natbib}
\usepackage[utf8]{inputenc}
\usepackage[english]{babel}
\usepackage{amsmath}
\usepackage{amsfonts}
\usepackage{amssymb}
\usepackage{graphicx}
\usepackage{epstopdf}
\usepackage{ulem}

\begin{document}
\title{Electronic correlation effects and Coulomb gap in the Si(111)-$(\sqrt{3}\times\sqrt{3})$-Sn surface}
\author{A.B. Odobescu}
\email[]{arty@cplire.ru}
\author{A.A. Maizlakh}
\author{N.I. Fedotov}
\author{S.V. Zaitsev-Zotov}
\affiliation{Kotel'nikov IRE RAS, Mokhovaya 11, bld.7, 125009 Moscow, Russia}

\date{\today}

\begin{abstract}
Electronic transport properties of the Si(111)-$(\sqrt{3}\times\sqrt{3})$-Sn surface formed on low doped Si substrates are studied using two-probe conductivity measurements and tunnelling spectroscopy. We demonstrate that the ground state corresponds to Mott-Hubbard insulator with a band gap $2\Delta = 70$~meV, which vanishes quickly upon temperature increase. The temperature dependence of the surface conductivity above $T > 50$~K corresponds to the Efros-Shklovskii hopping conduction law.  The energy gap at the Fermi level observed in tunnelling spectroscopy measurements at higher temperatures could be described in terms of dynamic Coulomb blockade  approximation. The obtained localization length of electron is $\xi = 7$~\AA. 
\end{abstract}
\maketitle

\section{Introduction}
Two-dimensional electron gas formed on reconstructed semiconductor surfaces where Coulomb interaction energy is comparable with charge carrier kinetic energy is very promising systems for experimental study the effect of electron-electron interactions on the ground state and temperature-dependent transport properties. Of great interest for the study are  $\alpha$-phase surfaces obtained by covering an ideal (111) Si or Ge surface with the 1/3 ML of group I, IV adatoms forming a $\sqrt{3}\times\sqrt{3}-R30^{\circ}$ reconstruction with $\sim 7$~\AA\  inter-adatom distance. These systems have a very narrow half-filled surface state band, with an odd number of electrons per unit cell, which is metallic but affected by various scenarios of structural and electronic instabilities at low temperatures: charge-density wave-like reversible $3 \times 3$ periodic surface structure distortion in Pb/Ge(111) and Sn/Ge(111) \citep{carpinelli, petersen}; undistorted $\sqrt{3}\times\sqrt{3}$ Mott-Hubbard insulator (MI) state in Si/SiC(0001)\citep{weitering} and K:Si(111)$-(\sqrt{3}\times\sqrt{3})R30^{\circ}$-B \citep{johansson}.

An intermediate case is the Si(111)-$(\sqrt{3}\times\sqrt{3})R30^\circ$-Sn surface.  For this structure a density functional calculation in the local density approximation (LDA) predicts an undistorted ground metal state \citep{profeta1}. The experimental study of the electronic surface structure by photoemission  shows a metallic band structure as expected from the simple electron counting but with clear indication of $3\times3$ periodicity \citep{uhrberg, lobo}. Structural tools like STM and photoelectron diffraction showing only $\sqrt{3}$ periodicity provide no fingerprints of the $3\times3$  structure  \citep{morikawa, lobo}. More recent local spin density (LSDA) + Hubbard U approximation \citep{profeta2} predicts a stable magnetic (Mott-Hubbard-like) insulating $\sqrt{3}\times\sqrt{3}$ phase in Sn/Si(111) which forms two separate $3\times 3$ unit cells where  one is fully ferromagnetic  and the other ferrimagnetic (two atoms spins up and one spin down). The calculation explains the fingerprints in photoemision intensity near $E_F$ at $\Gamma$ observed in Sn/Si(111) \citep{uhrberg} due to $3\times3$ anti-ferromagnetic order.  A scanning tunnelling spectroscopy (STS) demonstrated an insulating gap in the local density of states (LDOS) at the Fermi level \citep{modesti} with a narrow band gap ($2\Delta = 40$ meV), which disappears above 70K but the deep minima in LDOS still exists up to the room temperature. The answer on possible MI transition could be obtained from electronic transport measurements. Temperature dependence of conductivity of  the Sn/Si(111)$\sqrt{3}\times\sqrt{3}$ surface shows an insulator behaviour in a wide temperature range 20--300~K with possible variable-range hopping transport mechanism due to the defects or activation behaviour with very small activation energy $2\Delta \approx 10$ meV, which was not uniquely determined \citep{hirahara}.

Similar behaviour and discrepancy in various experimental results was observed in metallic Si(111)$7\times7$ surface \citep{losio, tanikawa, modesti2}. The recent explanation \citep{odobescu2} considers the narrow band gap observed in the surface states at the Fermi level as a result of the dynamic  Coulomb blocade which appears due to STS on surfaces with the sheet conductivity much less than the minimum metallic conductivity $1/R_h =e^2/h = 39 \mu S$ \citep{joyez}. The deep band gap at LDOS in Sn/Si(111)$\sqrt{3}\times\sqrt{3}$ surface behaves very similar to the dynamic Coulomb gap in Si(111)-$7\times 7$ and may follow the same scenario.      

Here we report the results of  detailed study of Si(111)-($\sqrt{3}\times\sqrt{3})$-Sn surface. We show by comparing the  temperature dependence of LDOS measured by tunnelling spectroscopy with temperature dependence of the surface conductivity,  that the surface ground state corresponds to insulator-like state with the surface band gap  $2\Delta = 70$ meV,  in accordance with LSDA+U calculation \citep{profeta2}. With increasing the temperature the MI spin-ordered state vanishes due to thermal fluctuation and the surface turns into a bad metal. The transport measurements indicate the Efros-Shlklovskii hoping regime with  the charge carrier localization length $\xi \approx 7$~ \AA, right to the  interatomic distance of Si(111)-($\sqrt{3}\times\sqrt{3})$-Sn unit cell, and the energy gap in tunnelling spectroscopy at higher temperatures is well described with temperature dependent dynamic  Coulomb gap approximation \citep{ joyez, brun}.    

\section{Experimental}
The tunnelling spectroscopy measurements were performed in commercial UHV LT STM Omicron. We used low doped {\it n-} and {\it p}-type Si crystals with $\rho = 1$~$\Omega\cdot$cm as the substrate in order  to eliminate the contribution of the bulk conductivity at low temperatures in electron transport measurements and the effect of charge carrier concentration change due to the band bending near the surface \citep{append1}.  Such crystals are insulating in the low-temperature region and cannot be studied by the  STM technique. Therefore, for STS study we used the external illumination to produce necessary bulk conduction at low temperatures (see details below). 

\begin{figure}
 \begin{tabular}{cc}
	(a) &\resizebox{60mm}{!}{\includegraphics{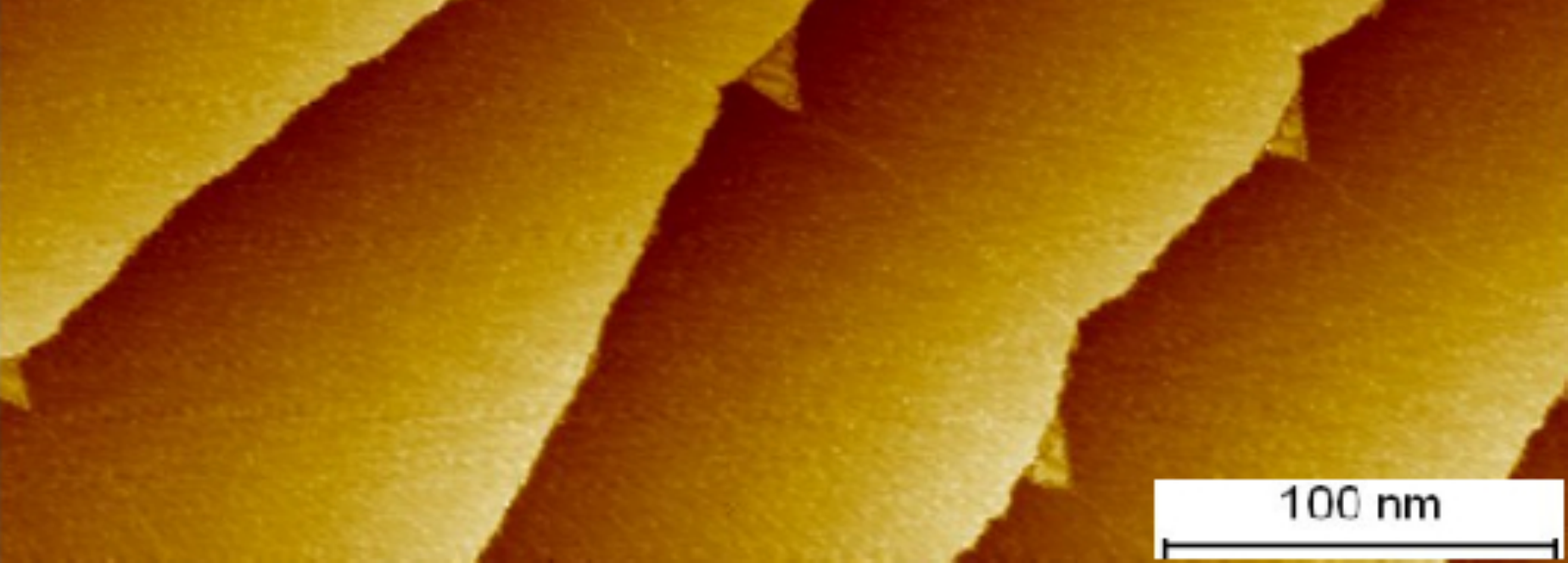}} \\
	(b) &\resizebox{60mm}{!}{\includegraphics{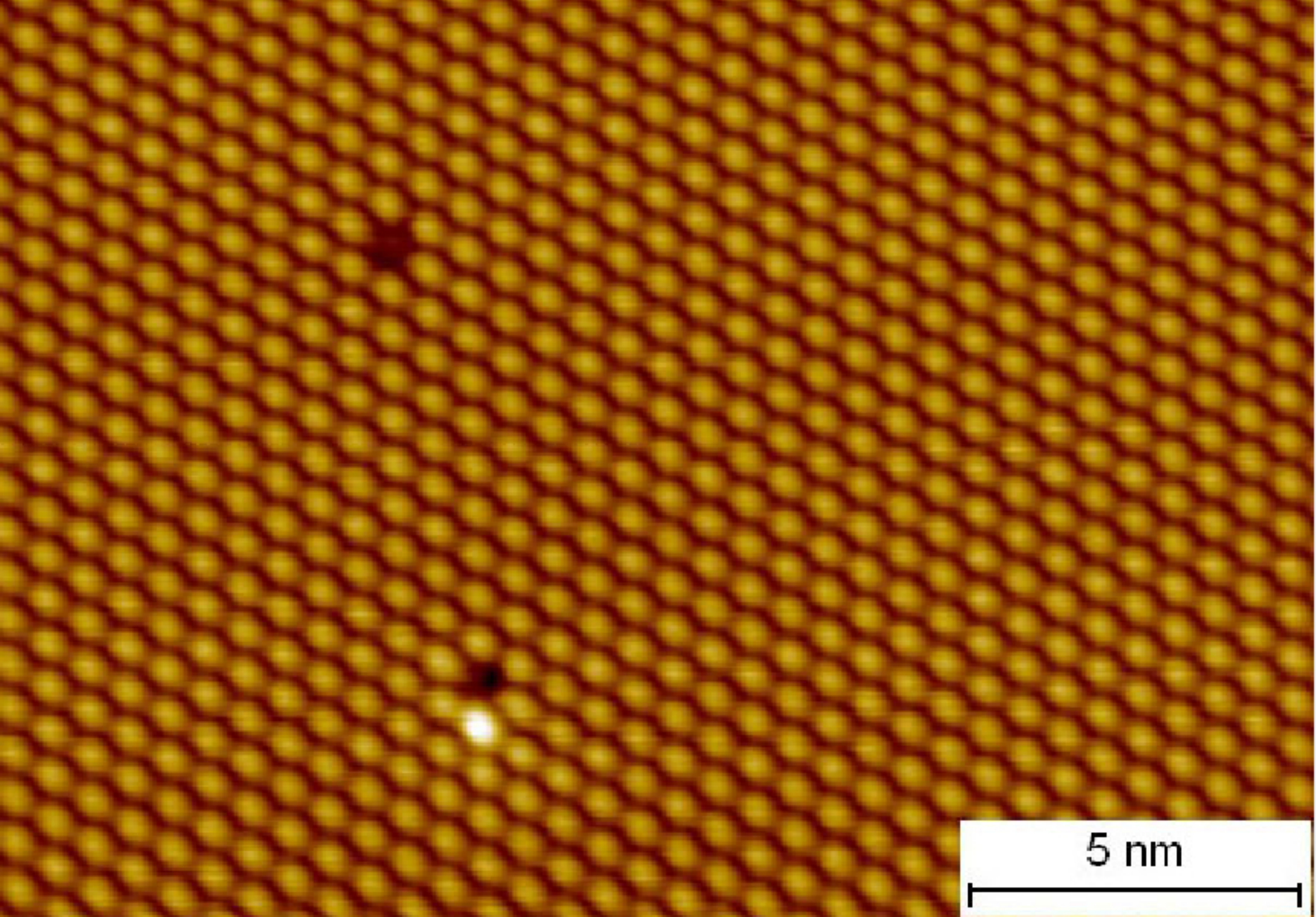}}
 \end{tabular}
\caption{(a) - the STM image of the Si(111)-($\sqrt{3}\times\sqrt{3}$)-Sn surface with different exposition time during Sn evaporation. (a): The extra Sn deposited adatoms forms triangle islands of the ($\sqrt{7}\times\sqrt{3}$)-Sn phase near terraces edges,  $T = 78$K; (b) - the clean Si(111)-($\sqrt{3}\times\sqrt{3}$)-Sn structure with average defects value less than $2\%$ is done at $T = 5$K under external illumination of the surface. Substrate Si \textit{n}-type, $\rho = 1$ $\Omega$cm,  $U_T = -2$ V, $I_T = 50$ pA.}
\label{stm_images}
\end{figure}

The Si(111)-$7\times7$ clean surface was prepared by direct current heating up to $1250^\circ$C for 30 s and computer controlled cooling. Then the 1/3 ML of Sn was deposited at RT on the surface and annealed at $600^\circ$C for 5 min to create the Si(111)-($\sqrt{3}\times\sqrt{3}$)-Sn reconstructions. To eliminate the ($\sqrt{7}\times\sqrt{3}$)-Sn phase growth near the step edges (Fig.~\ref{stm_images}a)  the necessary quantity of deposited Sn was fitted by reducing the evaporation time until the $7\times3$ - Sn phase disappears completely.
All STM and STS measurements were performed with platinum STM tips, preliminary tested on Au foil to verify spectroscopy and microscopy capability. The {\it I-V} curves were collected at fixed temperature over the adatom of the Si(111)-($\sqrt{3}\times\sqrt{3}$)-Sn unit cell far away from defects and averaged over series of measurements consisting of tens individual cycles. The data collected over the individual defects are the same. The {\it dI/dV} were measured with lock-in technique with a sinusoidal modulation of the sample bias voltage with amplitude 30-50 mV at frequency \textit{f} = 778 Hz, and verified by comparison with calculated numerically {\it dI/dV} from averaged {\it I-V} curves. 

The temperature dependent transport measurements were performed with two-probe technique using Omicron LT SPM with modified standard tip holder, equipped with two $\oslash30$~$\mu$m platinum wire contacts. The distance between point contacts was $l \approx 400$~$\mu$m. The {\it I-V} curves were measured using picoammeter Keithley 6487 with integrated voltage source at fixed sample temperatures controlled by LakeShore 325 temperature controller. 
In approximation of low conductive bulk the total conductivity accounts only 2D geometrical spreading resistance of the surface:
$$\sigma_{2p} = \frac{I}{\pi V}ln(\frac{l}{d})$$
where $d$ is the contact spot size. We take $ln(l/d)\approx 4.4$ which corresponds to $d=1-30$~$\mu$m within $30\%$ accuracy.
The contribution of the bulk conductivity into the measured data for used low doped samples was found to be negligibly small at low temperatures: a control measurement obtained for the surface degraded at vacuum condition during 7 days shown at Fig. \ref{sigma_T}. The conductivity of the bulk plus contribution through degraded surface $\sigma$ is more than 2 orders of magnitude smaller than the clean surface conductivity below $T = 100$~K . This gives the bulk contribution to be below $1\%$. The choice of two-probe technique let us to extend the measurements over 7 orders of surface conduction variation into a region of very low conductivity not available with usual 4-probe technique.

To perform spectroscopy measurements at helium temperatures we used external illumination from typical LED source with light intensity up to $10^{-4}$ W$\cdot$cm$^2$ at the sample position. This provides necessary conduction of the bulk required for STS and STM study and also removes bands bending near the surface and restores initial charge carrier concentration in the surface bands \citep{grafstrom, odobescu1}. Illumination-induced shift of the surface bands energy positions by photovoltage  (Fig. \ref{30K_sts}) was taken into account.

\section{Results}
\begin{figure}
    \resizebox{70mm}{!}{\includegraphics{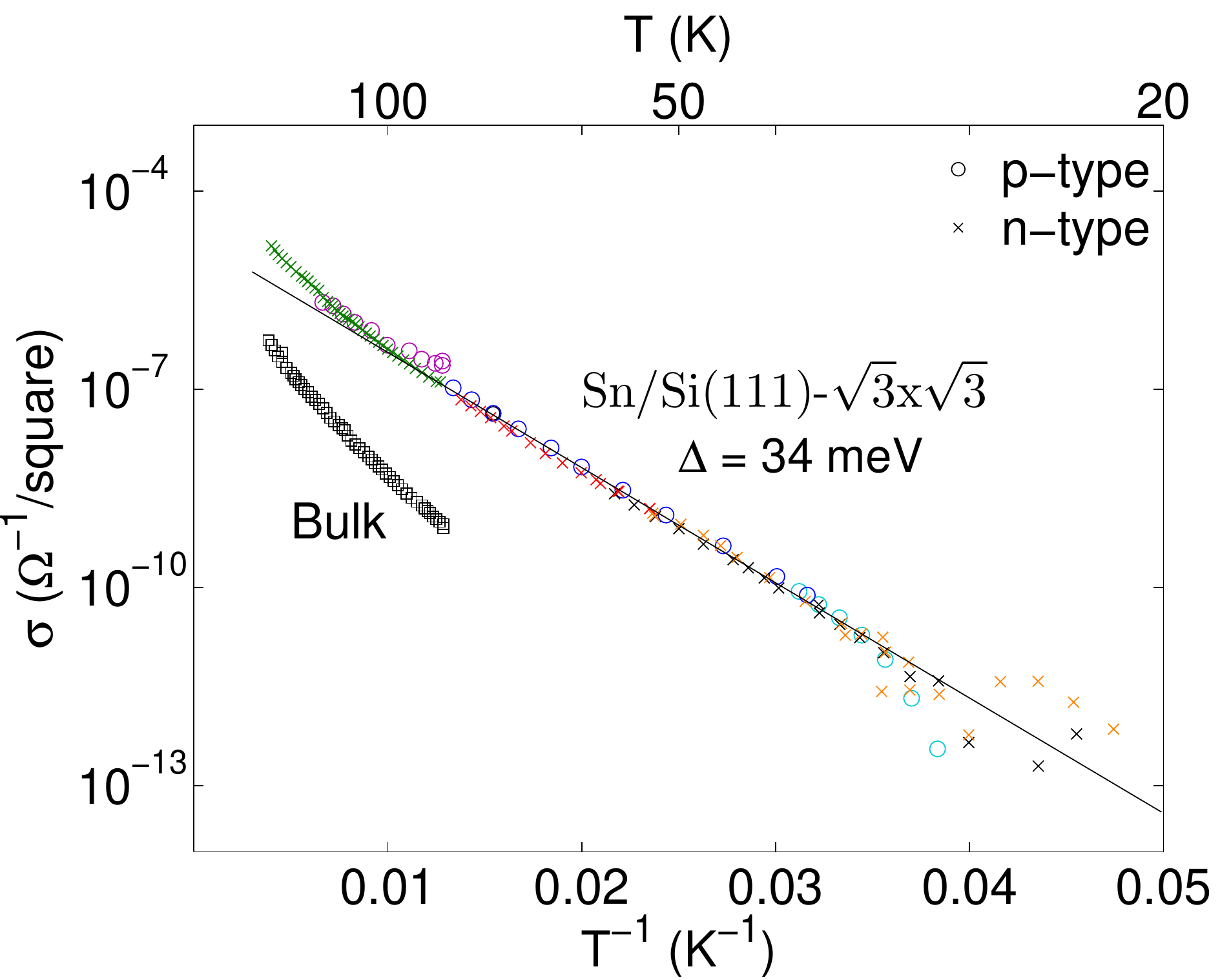}}
\caption{Temperature variation of the surface conductivity for the clean Si(111)-($\sqrt{3}\times\sqrt{3}$)-Sn structure obtained with various 2 micro probes on \textit{n-}type (crosses) and \textit{p-}type (circles) Si substrates with $\rho = 1$ $\Omega\cdot$cm. Different coloured markers correspond to the individual measurements done with various 2 micro probes on each time fresh prepared surface. The black squares are the 2-point measurements of the Si(111)-($\sqrt{3}\times\sqrt{3}$)-Sn surface degraded in the vacuum chamber for 7 days, indicates the contribution of the bulk and subsurface layer into the measured conductivity.}
\label{sigma_T}
\end{figure}

Fig.~\ref{sigma_T} shows the temperature dependence of the surface conductivity of the Si(111)-($\sqrt{3}\times\sqrt{3}$)-Sn surface measured on \textit{n-} and \textit{p-}type samples with various 2-point probes in the temperature range of $25-260$~K. At  $80 < T < 20$~K the conductivity corresponds to activation law $\sigma \propto \exp(-\frac{\Delta}{kT})$  with the activation energy $\Delta = 34$~meV, which is the same for both \textit{n-} and \textit{p-}type samples.  At lower temperatures ($T < 25$~K) the conductivity of the surface drops below the measurement limit of the used picoampermeter. The 25~K also was the limit for STM measurements on our samples without using of external illumination of our samples. 

\begin{figure}
 \begin{tabular}{cc}
    \resizebox{42mm}{!}{\includegraphics{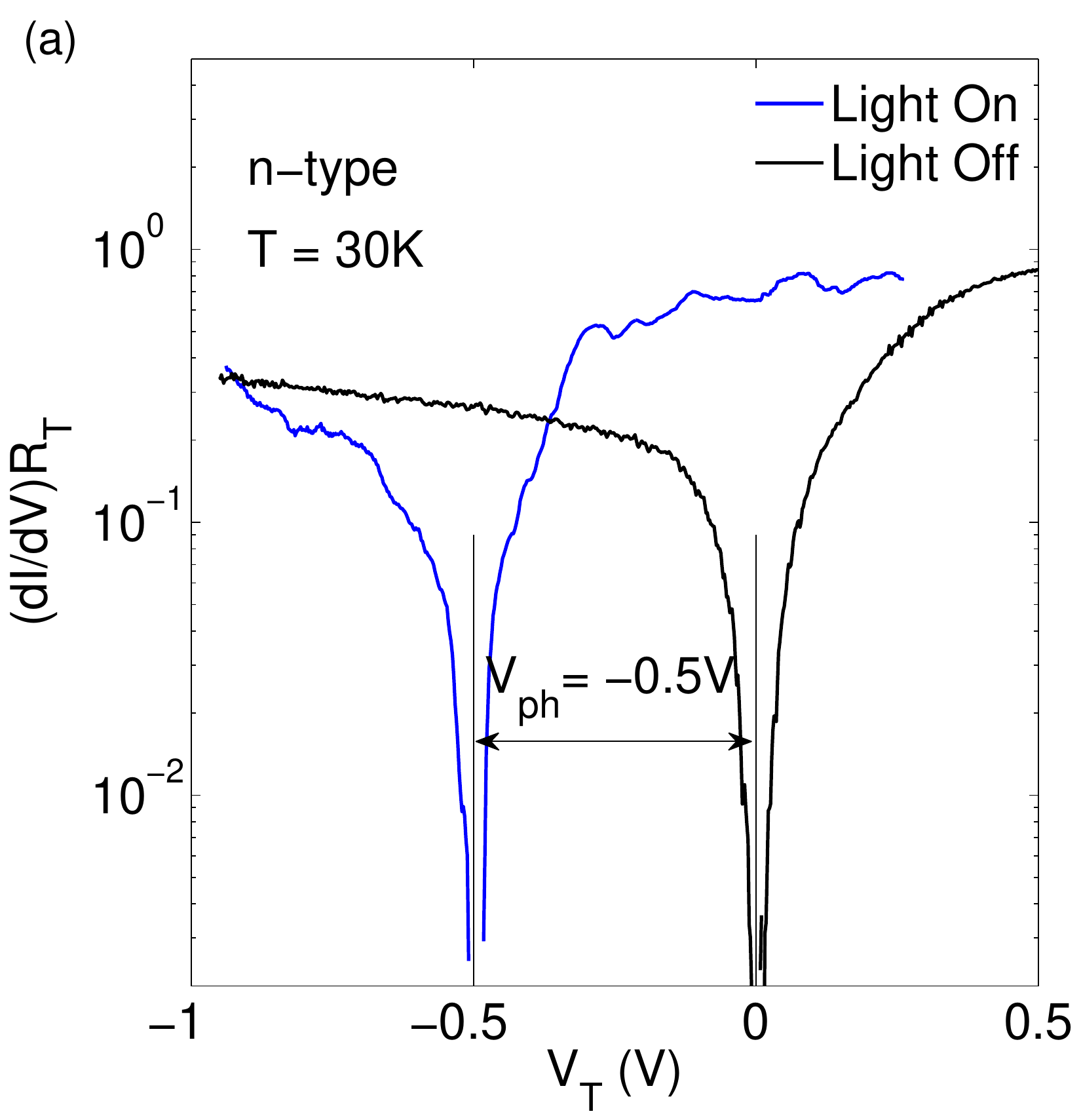}} & \resizebox{41mm}{!}{\includegraphics{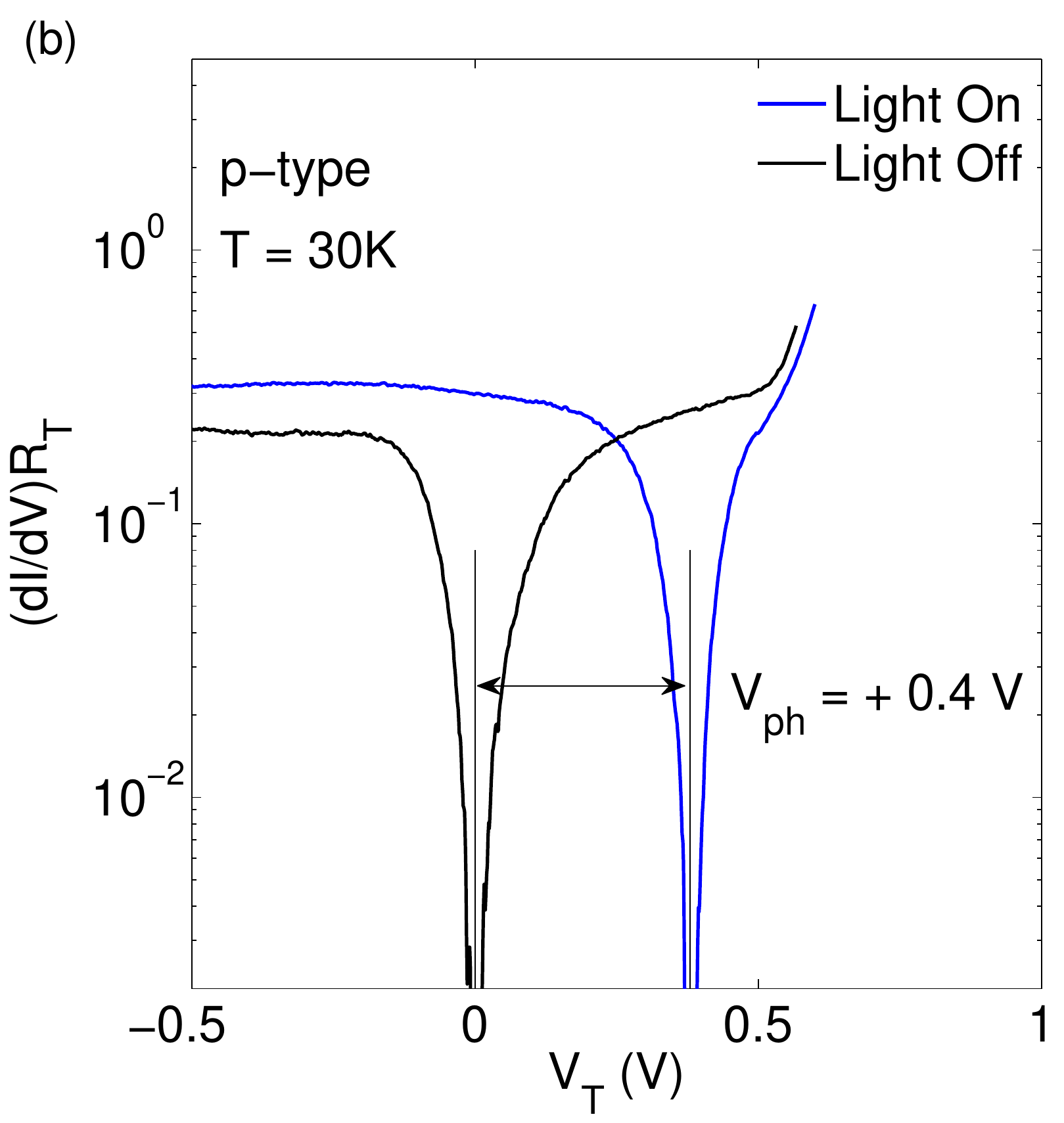}}\\
     \resizebox{41mm}{!}{\includegraphics{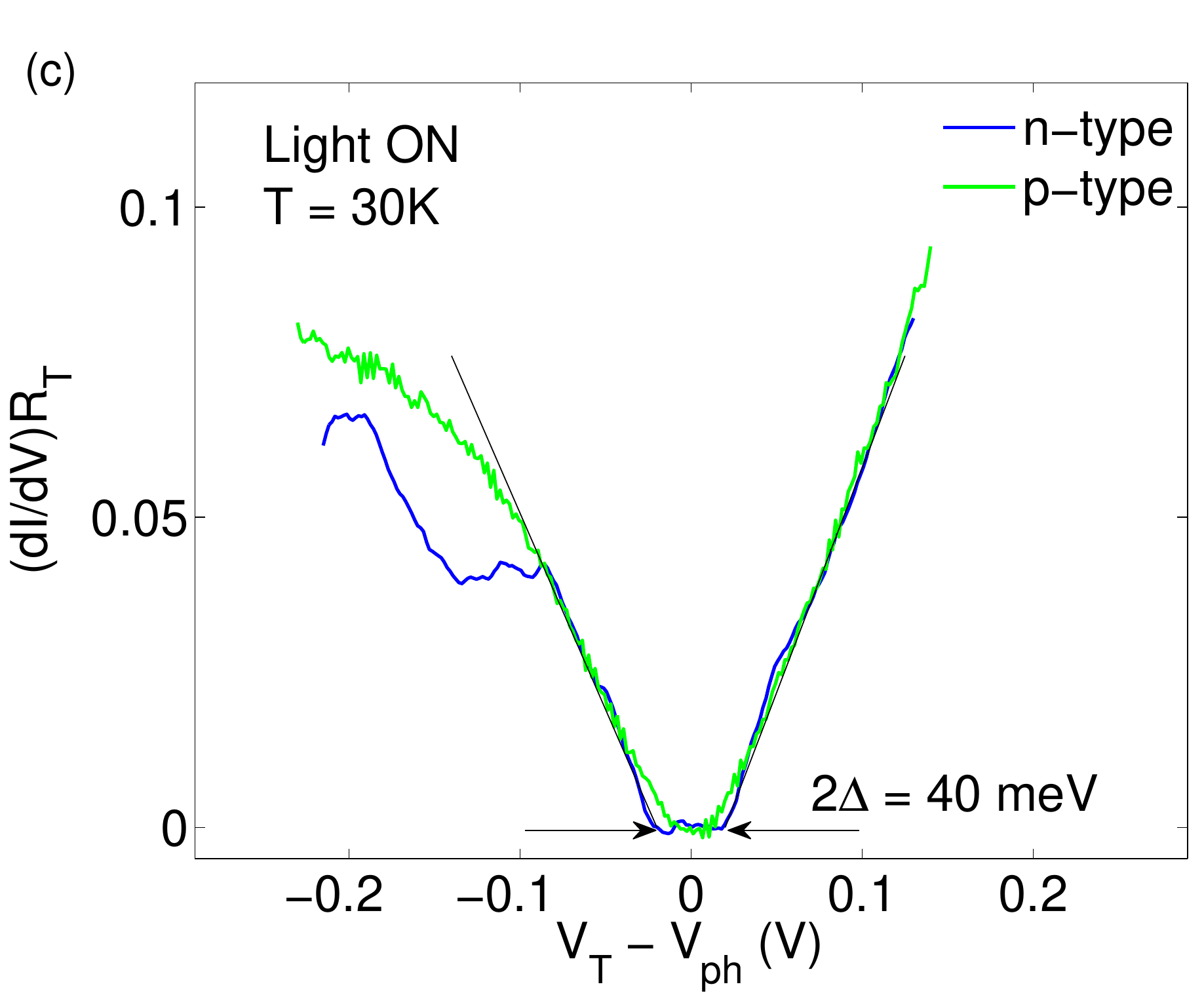}} & \resizebox{41mm}{!}{\includegraphics{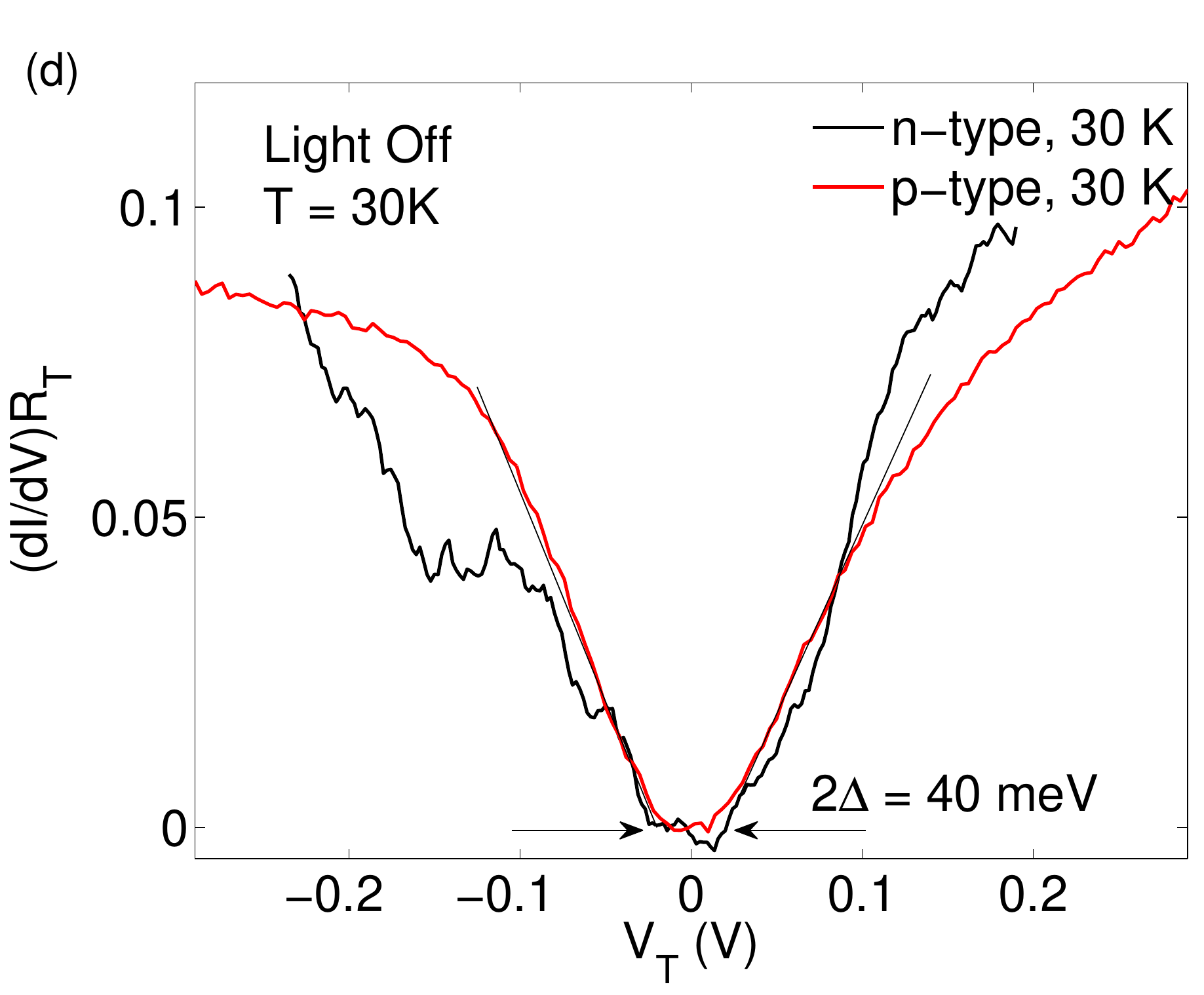}}\\
  \end{tabular}
\caption{(a-b): the STS spectra $(dI/dV)R_T$ of the Si(111)-($\sqrt{3}\times\sqrt{3}$)-Sn surface done at $ T = 30$K on illuminated (blue lines)\textit{n-} and \textit{p-}type samples, and the same point with no external illumination (black lines). (c-d): the value of the energy gap at Fermi level at  Si(111)-($\sqrt{3}\times\sqrt{3}$)-Sn surface for  \textit{n-} and \textit{p-}type samples depending on the illumination On or Off. Set point parameters are $U_T = -2$~V, $I_T = 50$~pA for \textit{n}-type, and  $U_T = 1.8$~V, $I_T = 50$~pA for \textit{p}-type.}
\label{30K_sts}
\end{figure}

Fig.~\ref{30K_sts} shows the STS data sets measured at $T = 30$K with and without external illumination. The energy gap at zero voltage is clearly seen on both data sets collected in the dark. When the samples are illuminated, the energy gap is shifted by the value of surface photovoltage $V_{ph} =  - 0.5$ V and $V_{ph} =0.4$ V for \textit{n-} and \textit{p-} samples respectively. The values of photoshifts indicate that the Shotky barrier near the surface almost completely disappears. We observed no noticeable deviation in the shape and value of the energy gap measured with and without external illumination (Fig. \ref{30K_sts}c-d). The value of the gap at $T = 30$~K is $2\Delta \approx 40$~meV  similar with reported in \citep{modesti} obtained on heavily doped Si samples at $T = 5$~K. 

\begin{figure}
 \begin{tabular}{cc}
    \resizebox{50mm}{!}{\includegraphics{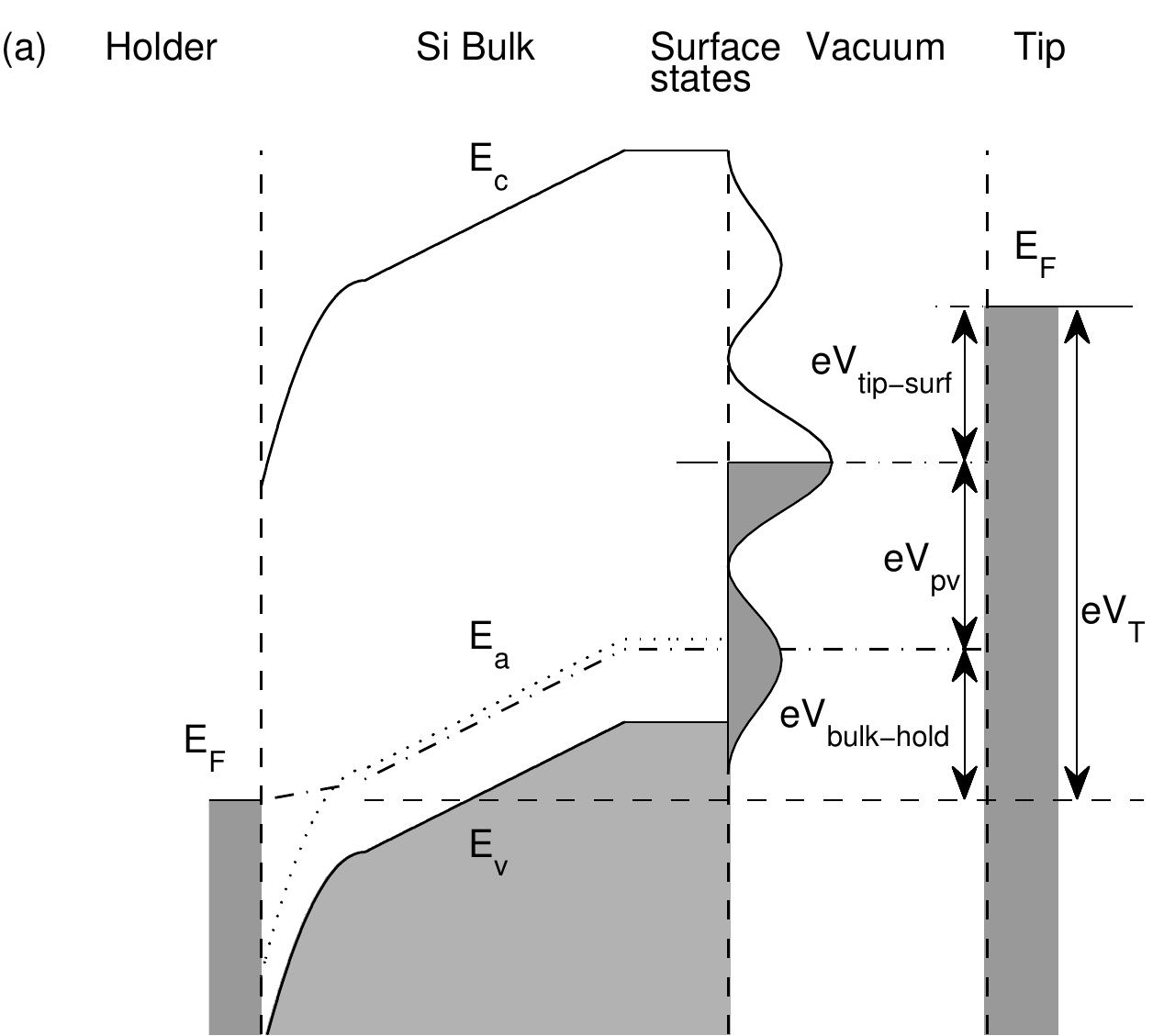}} & \resizebox{30mm}{!}{\includegraphics{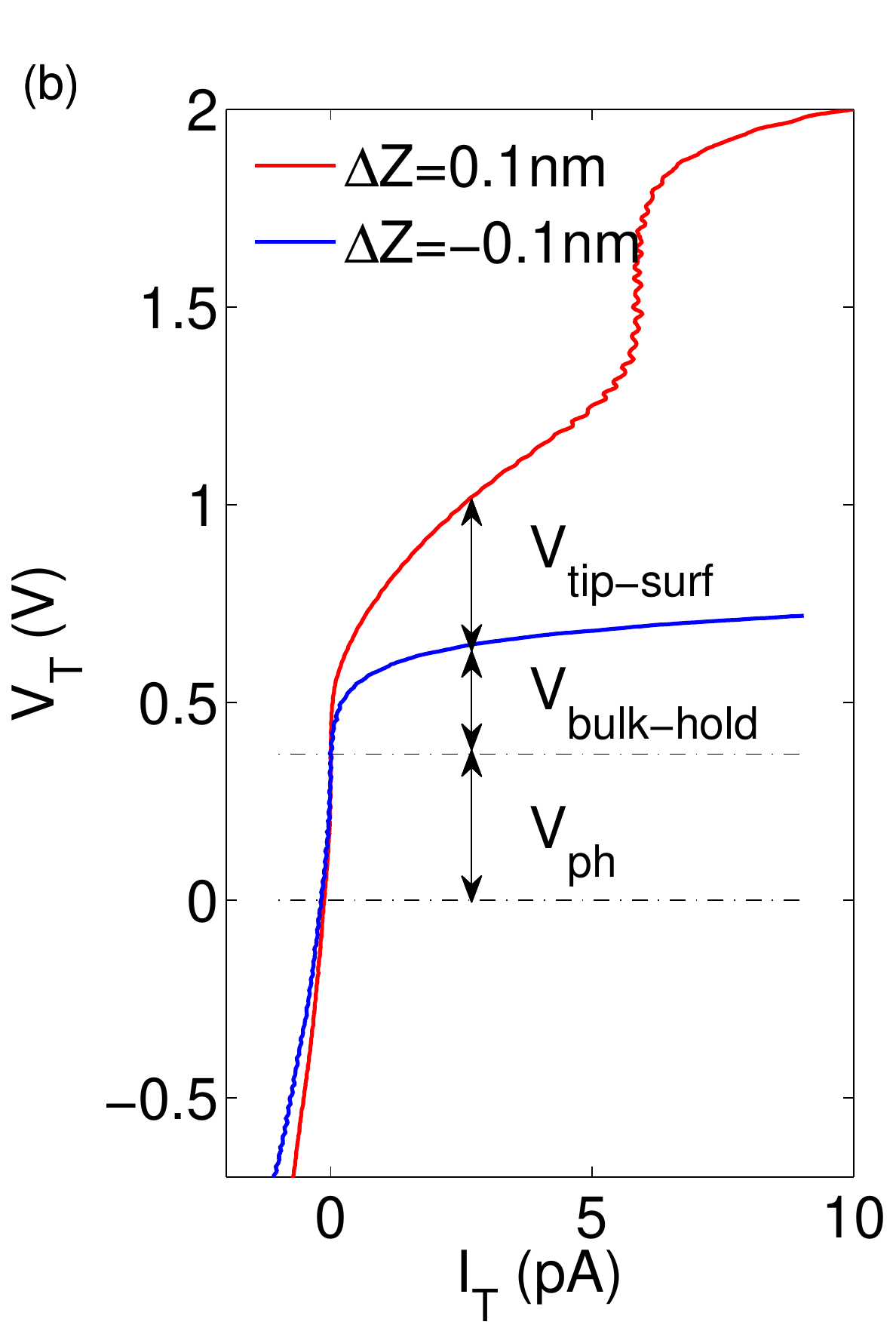}} \\
    \resizebox{30mm}{!}{\includegraphics{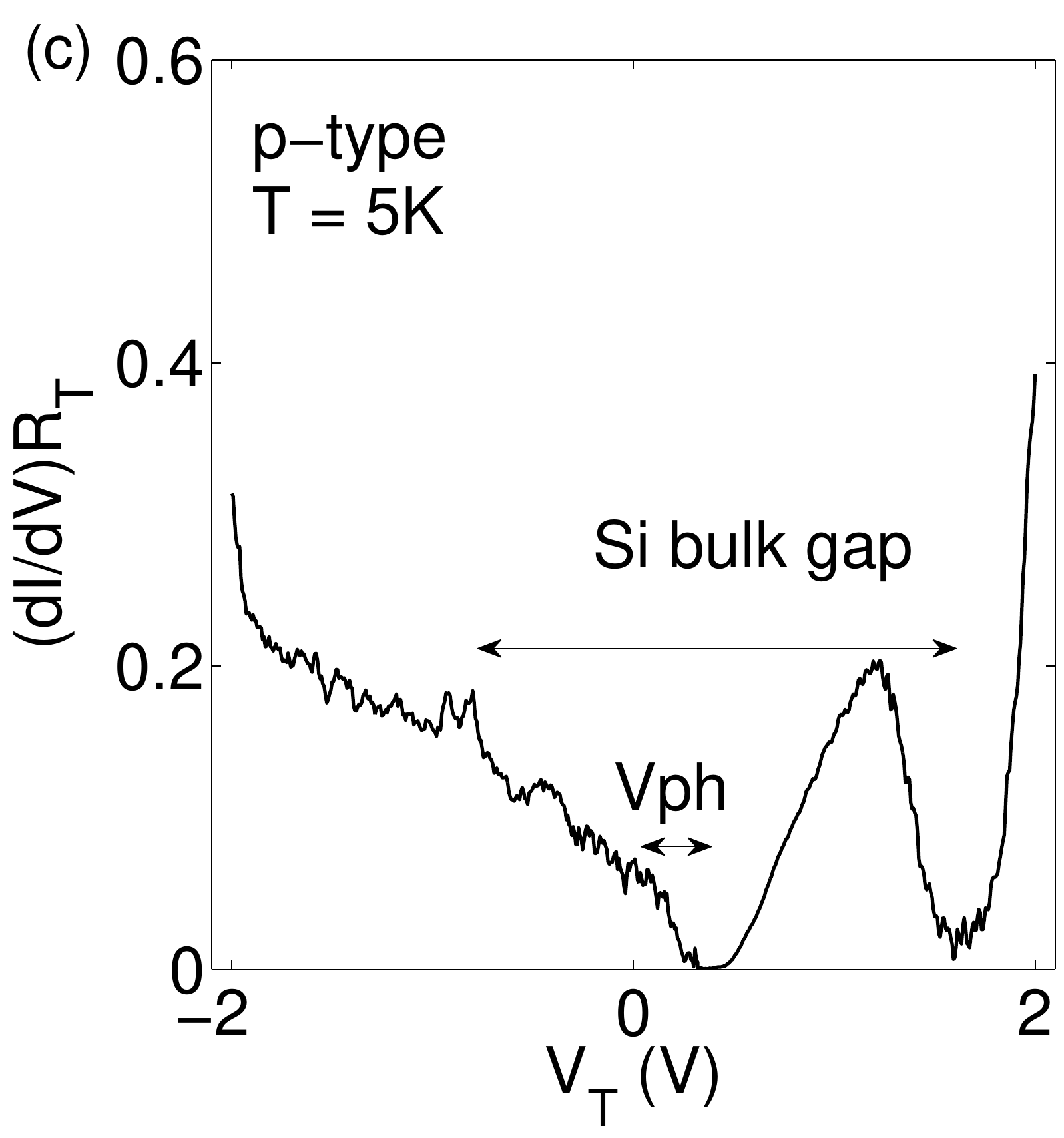}} & \resizebox{32mm}{!}{\includegraphics{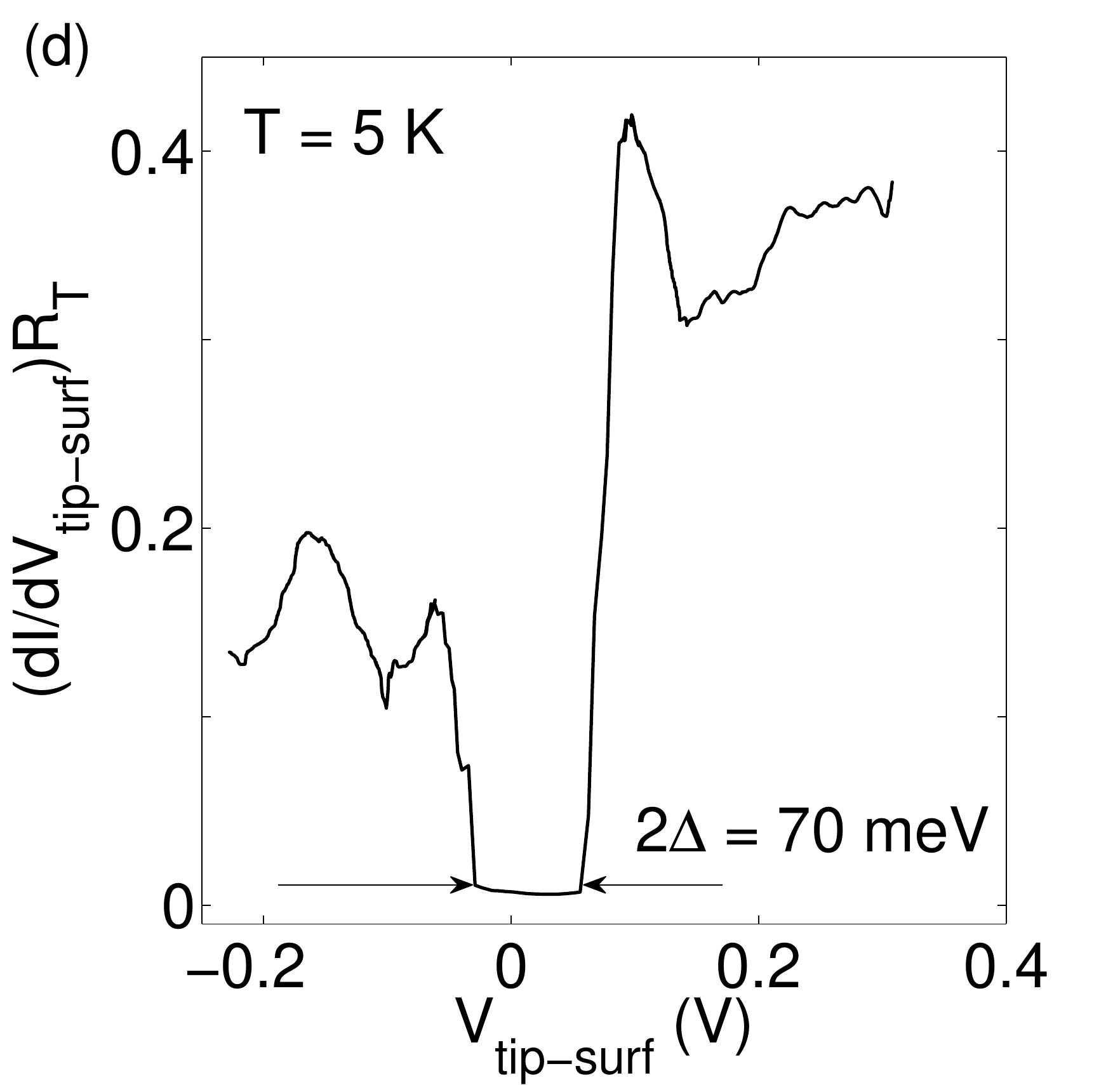}}
   \end{tabular}
\caption{(a) - schematic energy diagram of tip-surface-sample holder bias voltage distribution of the illuminated sample, when the resistance of the bulk-holder part is comparable with vacuum junction resistance; (c) - normalized $dI/dV$ data taken at set point $\Delta Z = 0$~\AA, $I_T = 50$~pA, $U_T = 2$~V; (b) - I/V data of the same point taken at different tip-sample distances shifted from set point at $\Delta Z$.  (d) - corrected normalized conductance $(dI/dV_{tip-surface})R_T$ of  the Sn/Si(111) surface at T = 5K.}
\label{en_diag}
\end{figure}

 At lower temperatures it was impossible to detect any surface states inside the bulk gap. The typical characteristic $dI/dV$ spectra of the bulk are plotted in Fig.~\ref{en_diag}c. The explanation for this is very low surface conductivity at helium temperatures and possible high resistance at contact sample-holder $R_{bulk-hold}$, comparable with tunnelling resistance $R_{tun}$. In this case the bias voltage $V_T$ is distributed partially between the vacuum tunnelling junction and (bulk Si)-(metallic sample holder) junction (Fig~\ref{en_diag}a). By altering the distance between the tip and surface the $V_{bulk-hold}$ part could be extracted and used for correction of $dI/dV$ data. The method is described in details in \citep{modesti2}. Namely, we measured $I(V)$ spectra at distances shifted from set point at $I_T = 50$~pA, $U_T = 2$~V,  $R_{T} =  4\cdot10^{10}$~$\Omega$ at $\Delta Z = -1$~\AA\ and $\Delta Z = + 1$~\AA\ (Fig~\ref{en_diag}b). Then $I(V_{tip-surface})$ is calculated the  by subtraction $I(V_{bulk-hold})$ from $I(V_T)$  measured at higher  tunnelling distances (Fig~\ref{en_diag}d). 
Then the energy gap value varying from 70 to 100 meV for both \textit{n-} and \textit{p-}type substrates was obtained from numerically  calculated $(dI/dV_{tip-surface})R_T$ spectra.

\begin{figure}
 \begin{tabular}{cc}
   \resizebox{35mm}{!}{\includegraphics{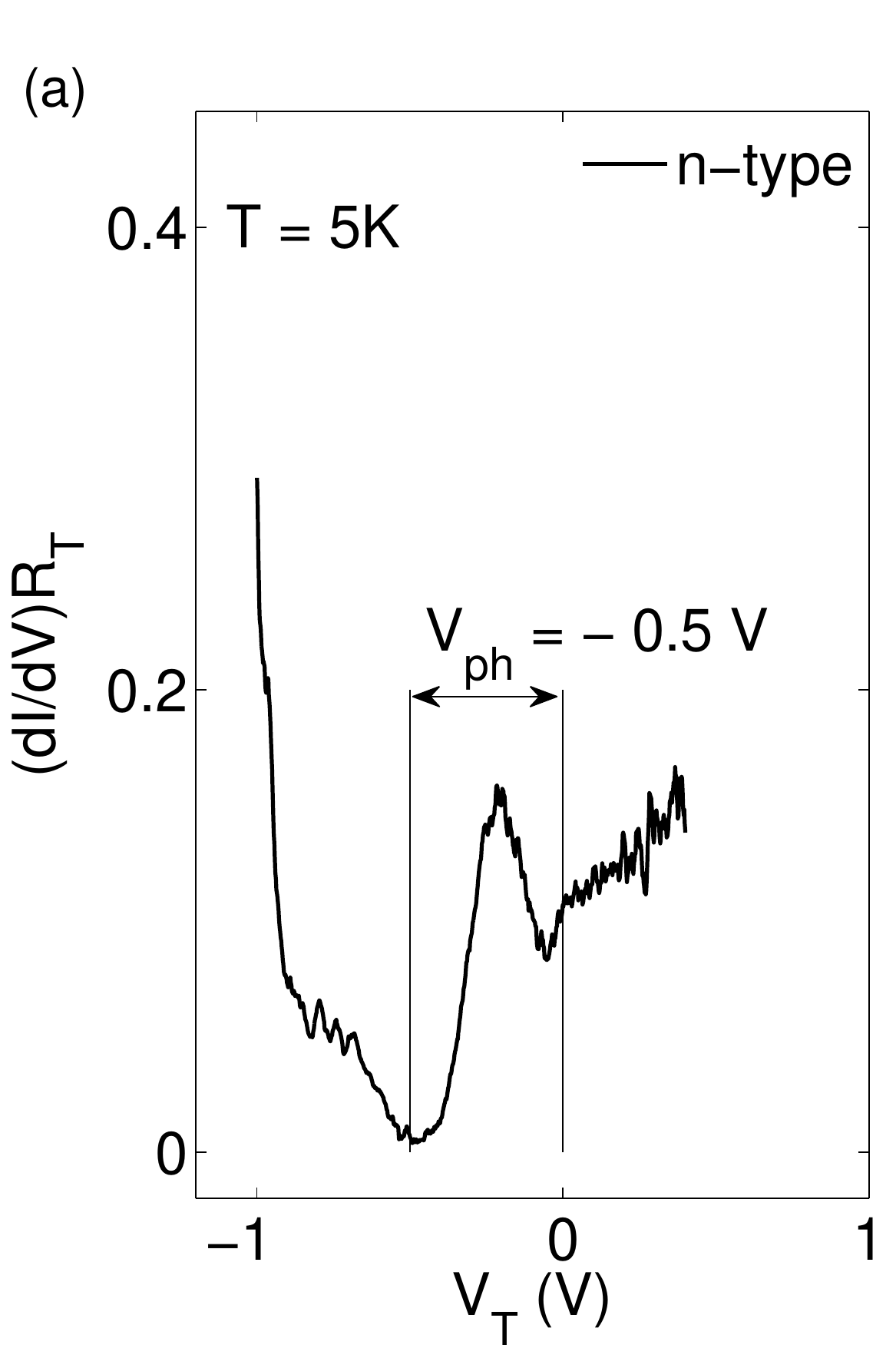}}  & \resizebox{35mm}{!}{\includegraphics{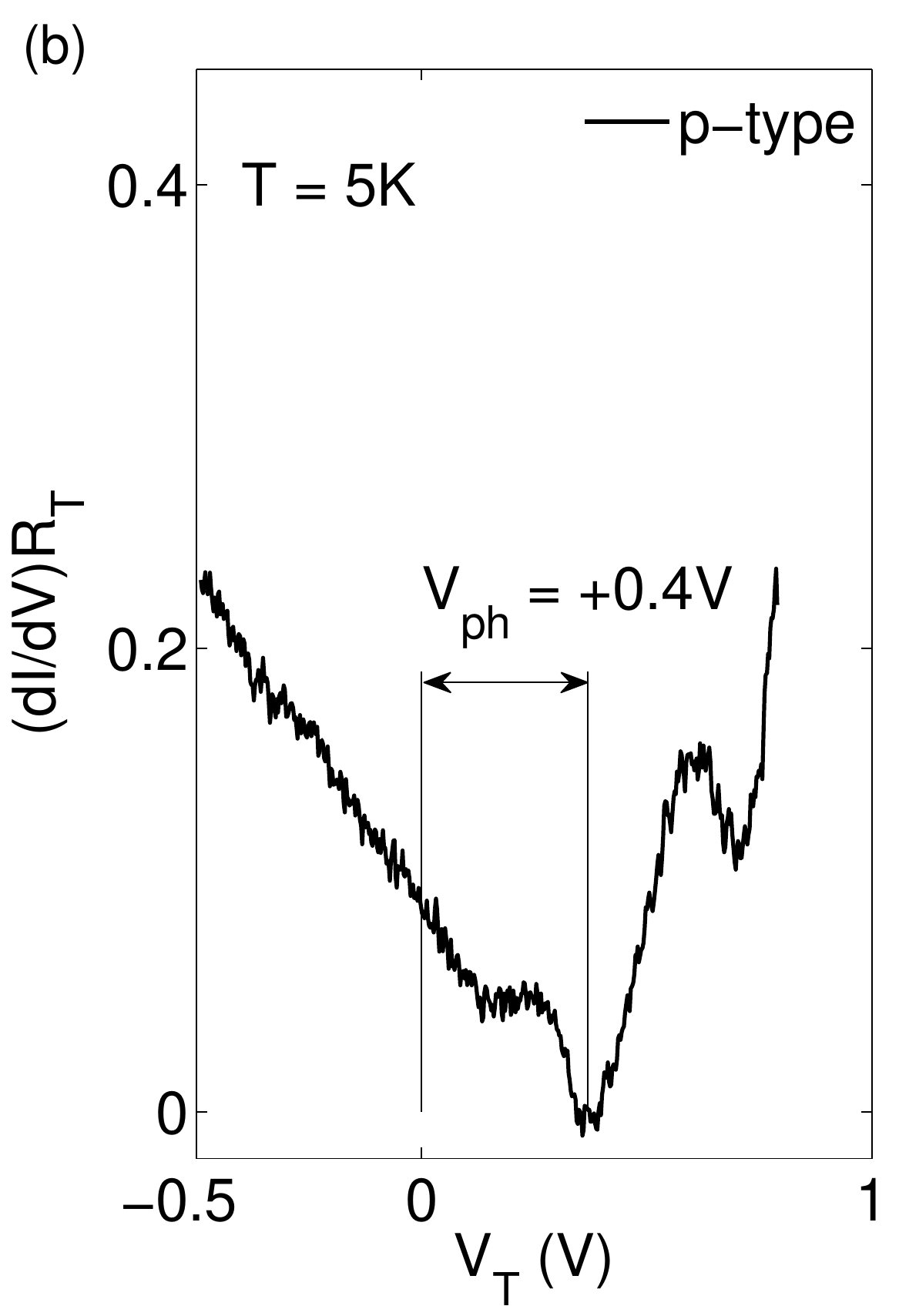}} \\
\end{tabular}
\caption{The normalized $(dI/dV)R_T$ data of the illuminated  Si(111)-($\sqrt{3}\times\sqrt{3}$)-Sn surface with preliminary deposited Ta contact areas, T = 5K. (a) - \textit{n}-type, (b) - \textit{p}-type Si substrate with $\rho = 1$ $\Omega$ cm. $U_T = -2$~V, $I_T = 50$~pA for \textit{n}-type, and  $U_T = 1.8$~V, $I_T = 50$~pA for \textit{p}-type. }
\label{5K_sts}
\end{figure}

Another method is to decrease the bulk-holder resistance by preliminary evaporation of Ta contact areas on the studied sample surface, so that the holder plates mount to this this Ta areas. In this case the $I(V)$ spectra was found to be not affected by the tip-surface distance, and the surface states could be distinguished in the bulk gap region. The resulted normalized $dI/dV$ spectra are shown on Fig.~\ref{5K_sts}. The  energy gap  values obtained by this method $2\Delta \approx 70$ meV for \textit{p}-type  sample and $2\Delta \approx 90$ meV for \textit{n}-type sample.

\section{Discussion}
\begin{figure}
\begin{tabular}{c}
   \resizebox{70mm}{!}{\includegraphics{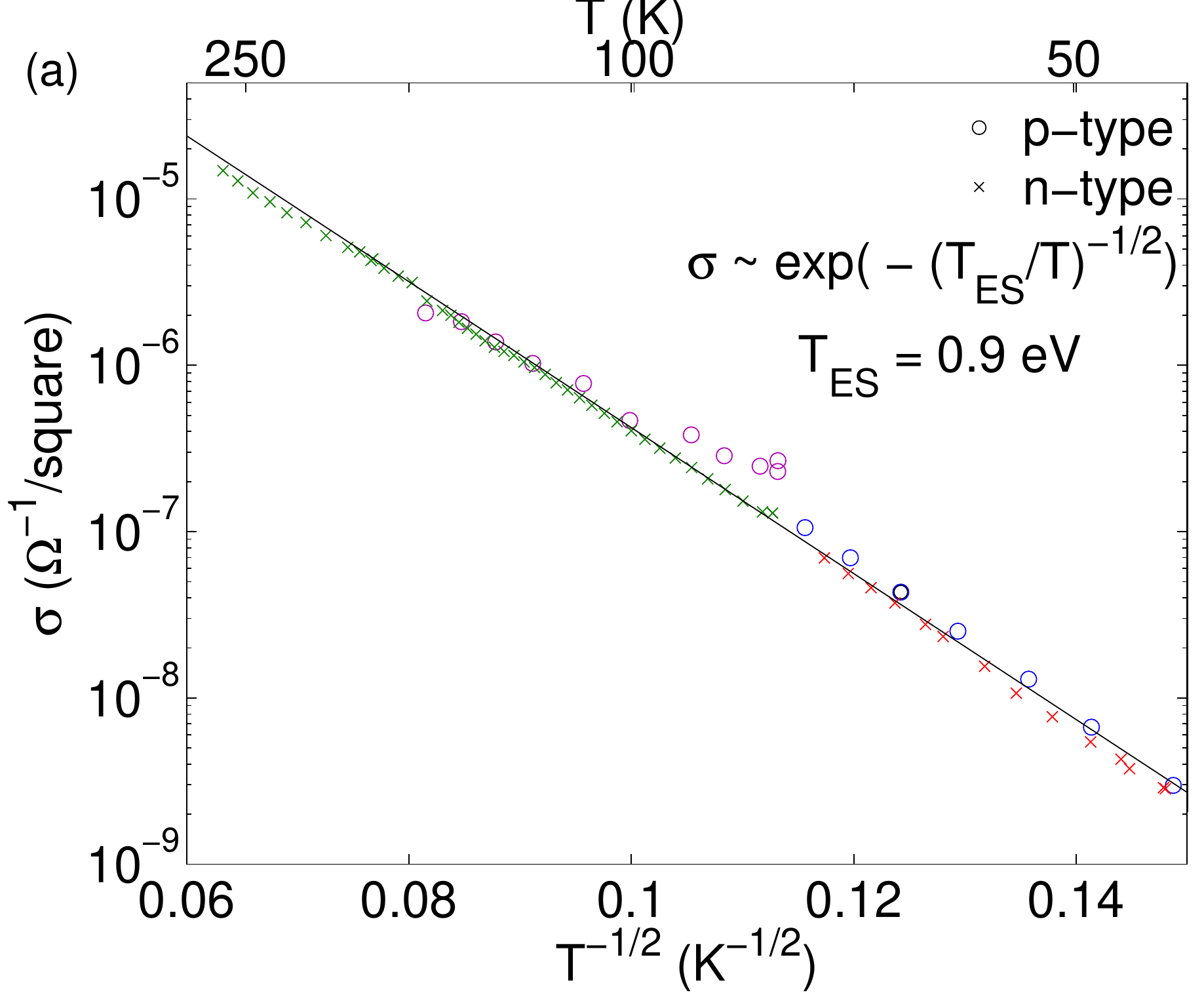}} \\
    \resizebox{70mm}{!}{\includegraphics{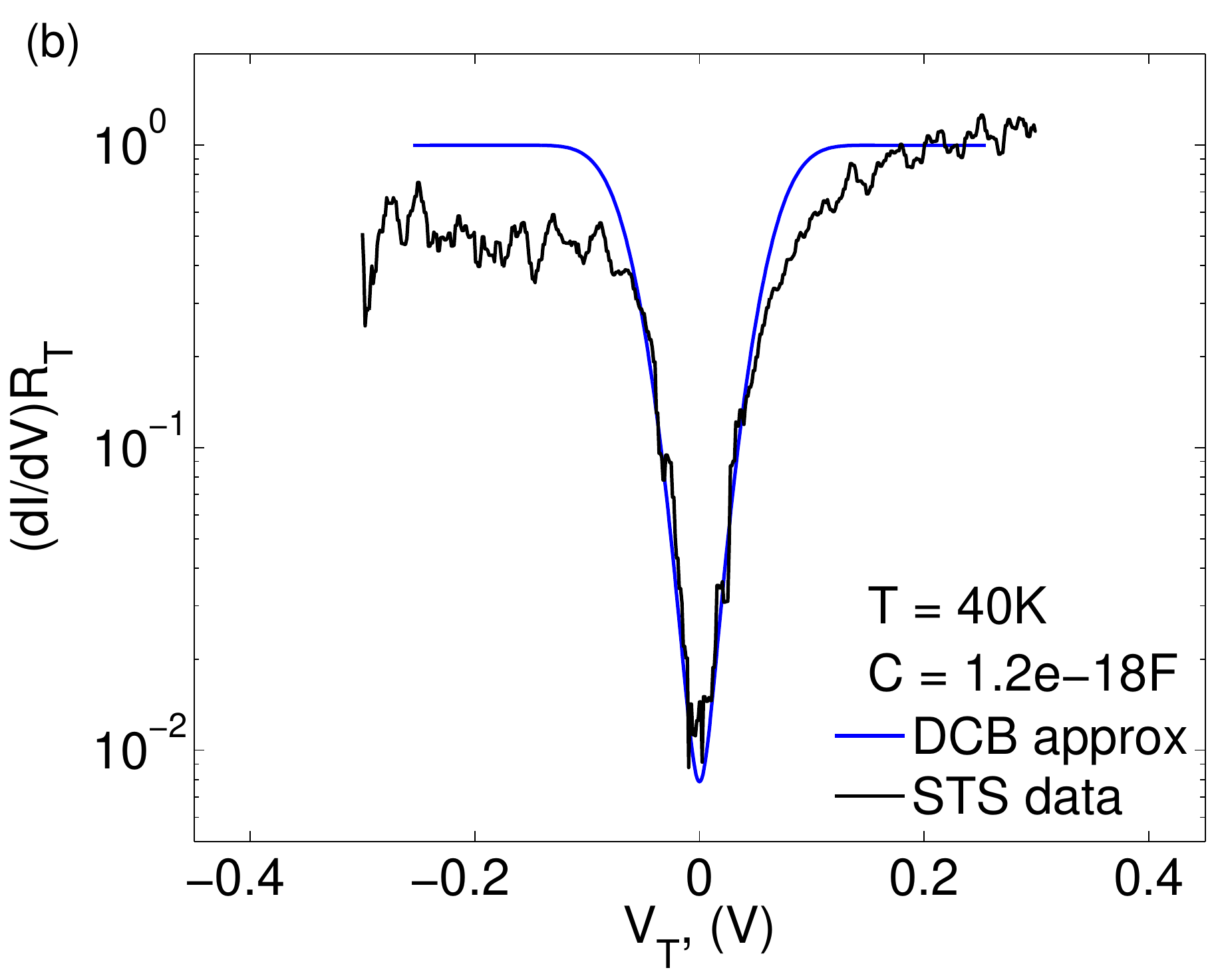}} 
\end{tabular}
\caption{(a) -the temperature variation of the Si(111)-$\sqrt{3}\times\sqrt{3}$-Sn surface conductivity presented in $T^{-1/2}$ coordinates. The data corresponds to the Efros-Sklovskii hopping conduction; (b) - the deep valley at surface Fermi level in normalized $(dI/dV)R_T$ spectra of the Si(111)-$\sqrt{3}\times\sqrt{3}$-Sn surface at T = 40~K (black line)  fitted within dynamic Coulomb approximation with $C_{surf} = 1.2\cdot10^{-18}$~F, $R_{surf} = 4\cdot10^{4}\cdot R_h$, $T = 40$~K. $U_T = -1.8$~V, $I_T = 50$~pA, sample \textit{n}-type Si. }
\label{EfSh}
\end{figure}

The measured value of the energy gap in LDOS for the  Si(111)-$\sqrt{3}\times\sqrt{3}$-Sn surface at helium temperature is $2\Delta \approx 70 \div 90$meV and agrees well to the activation energy obtained in the transport measurements $\Delta = 34$~meV. So the ground state corresponds to the calculated  spin-order magnetic Mott-Hubbard insulator state \citep{profeta2}. With temperature increase the spin ordering should disappear due to thermal fluctuations. The intersite antiferromagnetic (AF)  exchange coupling energy  is $J \sim 3$~meV \citep{profeta2}, so vanishing  of AF state at $T\gtrsim J\approx 40$~K is expected. The next phase for Si(111)-($\sqrt{3}\times\sqrt{3}$)-Sn surface is pseudo-gap metal. The conductivity is expected to be the variable-range hopping conduction affected by strong Coulomb interaction. Indeed, our data indicates the Efros-Shklovskii temperature dependent conductivity $\sigma \sim \exp(-(T_{ES}/T)^{1/2})$ above $T > 50$~K (Fig.~\ref{EfSh}a) with estimated $T_{ES} \approx 0.9$~eV in good agreement with our suggestion. This corresponds to the localization length of the charge carrier $\xi = \frac{2.8e^2}{\kappa T_{ES}}= 0.7$~nm \citep{shklovskii}, where $\kappa = (11.8 + 1)/2$ dielectric constant of the Si bulk plus vacuum. The localization length precisely matches the distance between Sn atoms in unit cell, which is very similar to what was observed in Si(111)-$7\times7$ surface \citep{odobescu2}.

With increasing the temperature the deep valley at zero bias in tunnelling LDOS is not connected any more to the band gap of the MI ground state, but could be described in terms of dynamic Coulomb blockade approximation as a sequential single electron tunnelling into a low conductive surface with sheet conduction $\sigma \ll e^2/h$. In this approximation the surface is presented as the resistance $R_{surf}$ and the capacitance $C_{surf}$  (for more details see \citep{joyez, brun}). The STS data at $T = 40$~K fitted within dynamic Coulomb approximation using the only one fitting parameter $C_{surf}$ is presented in Fig.\ref{EfSh}b. For higher temperatures the data also fits well within $C_{surf} = 1.1\cdot10^{-18}\div1.2\cdot10^{-18}$~F.  The relevant localization length $a$ of the tunnelled electron could be estimated as $a \sim C/8\pi \kappa \varepsilon_0 \approx 0.8$~nm \citep{append2}. The value agrees perfectly with the localization length obtained from the transport measurements.

\section{Conclusion}
In conclusion, through analysis of transport conductance measurements and tunnelling spectroscopy data of the Si(111)-$(\sqrt{3}\times\sqrt{3})$-Sn surface formed on low doped Si samples, we found that the ground state of this surface corresponds to the Mott-Hubbard insulator as was proposed before, with the surface band gap $2\Delta = 70$ meV. With increasing the temperature the magnetic insulator state vanishes due to thermal fluctuation and the surface turns into "bad" metal with a Coulomb gap at Fermi level.  The temperature dependence of the surface conductance corresponds to the Efros-Shklovskii conduction  at $T > 50$~K and the deep valley in STS data at higher temperatures  is described in terms of dynamic Coulomb approximation with measured localisation length perfectly agrees with interatomic Sn atoms distance.

\section{Acknowledgments}
\begin{acknowledgments}
The work was supported by the Russian Foundation for Basic Research  under contract 16-32-60156. 
\end{acknowledgments}


\begin{thebibliography}{1}
\bibitem{carpinelli}J.M. Carpinelli, H.H. Weitering, M. Bartkowiak, R. Stumpf, and E.W. Plummer, Phys. Reb. Lett. {\bf 79}, 2859 (1997).
\bibitem{petersen}L. Petersen, Ismail, and E.W. Plummer, Prog. Surf. Sci., {\bf 71} 1 (2002). 
\bibitem{weitering}H.H. Weitering et. al., Phys. Rev. Lett. {\bf 78}, 1331 (1997).
\bibitem{johansson}L.I. Johansson et. al., Surf. Sci. {\bf 360} L478 (1996).
\bibitem{profeta1}G.Profeta et. al., Phys. Rev. B. {\bf 62}, 1556 (2000).
\bibitem{uhrberg}R.I.G. Uhrberg, H.M. Zhang, T. Balasubramanian, S.T. Jemander, N. Lin, and G. V. Hansson, Phys. Rev . B. {\bf 62}, 8082 (2000).
\bibitem{lobo}J. Lobo, A. Tejeda, A. Mugarza, and E. G. Michel, Phys. Rev. B. {\bf 68}, 235332 (2003).
\bibitem{morikawa}H. Morikawa, I. Matsuda, and S. Hasegawa, Phys. Rev. B. {\bf 65}, 201308 (2002).
\bibitem{profeta2}G. Profeta and E. Tosatti, Phys. Rev. Lett. {\bf 98}, 086401 (2007).
\bibitem{modesti}S. Modesti, L. Petaccia, G. Ceballos, I. Vobornik, G. Panaccione, G. Rossi, L. Ottaviano, R. Larciprete, S. Lizzit, and A. Goldoni, Phys. Rev. Lett. {\bf 98},126401 (2007).
\bibitem{hirahara}T. Hirahara, T. Komorida, Y. Gu, F. Nakamura, H. Idzuchi, H. Morikawa, and S. Hasegawa, Phys. Rev. B. {\bf 80}, 235419 (2009).
\bibitem{losio}R. Losio, K. N. Altmann, and F. J. Himpsel, Phys. Rev. B {\bf 61}, 10845 (2000).
\bibitem{modesti2}S. Modesti, H. Gutzmann, J. Wiebe, and R. Wiesendanger, Phys. Rev. B {\bf 80}, 125326 (2009). 
\bibitem{tanikawa}T. Tanikawa, K. Yoo, I. Matsuda, S. Hasegawa, and Y. Hasegawa, Phys. Rev. B {\bf 68}, 113303 (2003).
\bibitem{odobescu2}A.B. Odobescu, A. A. Maizlakh and S. V. Zaitsev-Zotov, Phys. Rev. B \textbf{92}, 165313 (2015).
\bibitem{joyez}P. Joyez and D. Esteve, Phys. Rev. B {\bf 56} 1848 (1997).
\bibitem{brun}C.Brun, K.H. Muler, I. Hong, F. Patthey, C. Flindt, W.- D. Shneider, Phys. Rev. Lett. {\bf 108}, 126802 (2012).
\bibitem{append1}Current carrier concentration change $\Delta Q = d \cdot N_a$, where $d = \sqrt{ \dfrac{2 \epsilon \varepsilon \Delta E}{e^2 N_a}}$ - is the depletion depth, and $N_a$ - doppant concetration in the substrate, $\Delta E  \approx 0.4$ eV - band bending value. For heavy doped Si with $\rho = 0.001 \Omega\cdot$cm the carrier concentration change in $\sqrt{3}\times\sqrt{3}$-Sn surface is estimated about 10\%.
\bibitem{odobescu1}A.B. Odobescu and S.V. Zaitsev-Zotov, J. Phys.: Condens. Matter {\bf 24}, 395003 (2012).  
\bibitem{grafstrom}S. Grafstr{\"o}m, J. Appl. Phys. {\bf 91}, 1717 {2002}.
\bibitem{myslivecek}J. Myslive{\v c}ek, A. Str{\' o}{\. z}ecka, J. Steffl, P. Sobot{\' i}k, I. O{\v s}t'{\' a}dal, and B. Voigtl{\" a}nder, Phys. Rev. B {\bf 73}, 161302 (2006).
\bibitem{shklovskii}B.I. Shklovskii, A.L. Efros. Electronic properties of doped semiconductors. Springer-Verlag (1984).
\bibitem{append2}The capacitance of tiny disk with characteristic dimension \textit{a} describe spreading area of a tunnelled electron before the next tunnelling event is estimated as $C  = 8a\pi \kappa \varepsilon_0$. 
\end{thebibliography}
\end{document}